\documentclass[prl,reprint,superscriptaddress,showpacs]{revtex4-1}

\usepackage{dcolumn}
\usepackage{multirow}
\usepackage{epsfig}
\usepackage[usenames]{color}
\usepackage[english]{babel}
\usepackage{relsize}
\usepackage[switch, modulo]{lineno}

\definecolor{grey}{rgb}{0.9,0.9,0.9}
\definecolor{black}{rgb}{0,0,0}

\newcommand{\be}{\begin{eqnarray}}
\newcommand{\ee}{\end{eqnarray}}
\newcommand{\bc}{\begin{center}}
\newcommand{\ec}{\end{center}}
\newcommand{\beq}{\begin{eqnarray}}
\newcommand{\eea}{\end{eqnarray}}

\begin{document}

\title{\boldmath Strong evidence for nucleon resonances near 1900\,MeV
}

\newcommand*{\HISKP}{Helmholtz--Institut f\"ur Strahlen--
                     und Kernphysik, Universit\"at Bonn, 53115 Bonn, Germany}
\newcommand*{\GATCHINA}{National Research Centre ``Kurchatov Institute'',
                     Petersburg Nuclear Physics Institute, Gatchina, 188300 Russia}
\newcommand*{\JLAB}{Thomas Jefferson National Accelerator Facility, Newport News, Virginia 23606, USA}
\newcommand*{\TUZLA}{University of Tuzla, Faculty of Natural Science and Mathematics,
 Univerzitetska 4,  75000 Tuzla, Bosnia and Herzegovina}
\newcommand*{\GLASGOW}{SUPA, School of Physics and Astronomy, University of Glasgow, Glasgow G12 8QQ, United Kingdom}
\newcommand*{\ZAGREB}{Rudjer Boskovic Institute, Bijenicka cesta 54,
                      P.O. Box 180, 10002 Zagreb, Croatia}

\affiliation{\HISKP}
\affiliation{\GATCHINA}
\affiliation{\JLAB}
\affiliation{\TUZLA}
\affiliation{\GLASGOW}
\affiliation{\ZAGREB}

\author{A.V.~Anisovich} \affiliation{\HISKP} \affiliation{\GATCHINA}
\author{V.~Burkert} \affiliation{\JLAB}
\author{M.~Had\v{z}imehmedovi\'{c}} \affiliation{\TUZLA}
\author{D.G.~Ireland} \affiliation{\GLASGOW}
\author{E.~Klempt} \affiliation{\HISKP} \affiliation{\JLAB}
\author{V.A.~Nikonov} \affiliation{\HISKP} \affiliation{\GATCHINA}
\author{R.~Omerovi\'{c} } \affiliation{\TUZLA}
\author{H. Osmanovi\'{c} } \affiliation{\TUZLA}
\author{A.V.~Sarantsev} \affiliation{\HISKP} \affiliation{\GATCHINA}
\author{J.~Stahov} \affiliation{\TUZLA}
\author{A.~\v{S}varc} \affiliation{\ZAGREB}
\author{U.~Thoma} \affiliation{\HISKP}

\begin{abstract}
Data on the reaction $\gamma p\to K^+\Lambda$ from the CLAS experiments are used to derive
the leading multipoles, $E_{0+}$, $M_{1-}$,  $E_{1+}$,  and $M_{1+}$, from the production threshold to
2180\,MeV in 24 slices of the invariant mass. The four multipoles are determined without any
constraints. The multipoles are fitted using a multichannel $L+P$ model which allows us to
search for singularities and to extract the positions of poles on the complex energy plane
in an almost model-independent method.
The multipoles are also used as additional constraints in an
energy-dependent analysis of a large body of pion and photo-induced reactions within the
Bonn-Gatchina (BnGa) partial wave analysis. The study confirms the existence of
poles due to nucleon
resonances with spin-parity $J^P = 1/2^-; 1/2^+$, and $3/2^+$ in the region at about 1.9\,GeV.
\end{abstract}
\keywords{baryon spectroscopy  \sep meson photoproduction \sep polarization observables \sep  Laurent+Pietarinen formalism}
\maketitle
``Three quarks for Muster Mark'' \cite{JJ:1939}. This sentence inspired Gell-Mann
\cite{Gell-Mann:1990} to call {\it quarks} the three constituents of nucleons, of protons or
neutrons. As a three-body system, the nucleon is expected to exhibit a large number
of excitation modes. The most comprehensive predictions of the resonance excitation
spectrum stem from quark-model calculations~\cite{Capstick:1986bm}-\cite{Giannini:2015zia};
this predicted spectrum is qualitatively confirmed by recent Lattice QCD
calculations~\cite{Edwards:2011jj}, even though the quark masses used lead to a
pion mass of 396\,MeV. The predicted resonances
may decay into a large variety of different decay modes. The most easily accessible was,
for a long time, the $\pi N$ decay of nucleon excitations by studying $\pi^\pm p$
elastic scattering and the $\pi^-p\to\pi^0n$ charge exchange reaction. A large amount of data
were analyzed by the groups at Karlsruhe-Helsinki (KH)~\cite{Hohler:1984ux},
Carnegie-Mellon (CM)~\cite{Cutkosky:1980rh} and at GWU \cite{Arndt:2006bf}. Real and
imaginary parts of partial waves amplitudes with defined spin and parity ($J^P$) were
extracted in slices of the $\pi N$ invariant mass,
and resonant contributions were identified. However, only a small fraction of the predicted energy
levels has been observed experimentally, and for some of them, the evidence for
their existence is {\it only fair} or even {\it poor} \cite{classify,Olive:2016xmw}.

The small number of observed excitations of
the nucleon, as compared to quark model calculations,  led to a number of speculations: Are nucleon resonances quark-diquark
oscillations with quasi-stable diquarks
\cite{Anselmino:1992vg,Kirchbach:2001de,Jaffe:2003sg,Jaffe:2004ph,Santopinto:2004hw})\,?
Are resonances generated by
meson-baryon interactions \cite{Dalitz:1961dv,Schneider:2006bd,Kaiser:1995cy,Lutz:2005ip,Zou:2007mk,Mai:2012wy},
and are quarks and gluons misleading as degrees of
freedom to interpret the excitation spectrum\,?
Does the mass-degeneracy of high-mass
baryon resonances with positive and negative-parity hadron resonances indicate the
onset of a new regime in which chiral symmetry is restored
\cite{Glozman:1999tk,Glozman:2003bt,Glozman:2007ek}\,? At low excitation
energy, chiral symmetry is strongly violated as indicated by the large mass gap
between the nucleon mass (with spin-parity $J^P=1/2^+$) and its chiral partner
$N(1535)$ with $J^P$\,=\,$1/2^-$. A more conventional interpretation assumes that
the {\it missing resonances} may have escaped detection due
to their small coupling to the $\pi N$ channel \cite{Koniuk:1979vw}. Some
evidence exists, however, that resonances in this mass region can be produced
by electromagnetic excitation and decay into $K^+\Lambda$ \cite{Olive:2016xmw}.
Thus, the photoproduction reaction $\gamma p\to K^+\Lambda$ bears the promise
of revealing the existence of resonances that are only weakly coupled
to $\pi N$. Fits to pion {\it and} photo-produced reactions have been performed by several
groups (BnGa~\cite{Anisovich:2015gia}, EBAC (KEK-Osaka-Argonne)~\cite{Kamano:2016bgm},
Gie\ss en~\cite{Shklyar:2012js}, J\"uBo~\cite{Ronchen:2015vfa}, MAID~\cite{Kashevarov:2016owq},
SAID~\cite{Workman:2012hx}, and others), and a number of resonances
has been reported~\cite{Anisovich:2011fc}. The resonances stem from energy-dependent
fits to the data.
The resonances and the background contributions in all partial waves need to be determined in
a single step. New data on $\gamma p\to K^+\Lambda$ enable a reconstruction of the photoproduction multipoles as functions
of energy. The multipoles drive the excitation of one partial wave; hence the fits
need to determine only resonances -- and the background -- contributing to a single partial wave.

The photoproduction of pseudoscalar mesons with an octet baryon in the final
state is governed by four complex (CGLN) amplitudes
${\mathcal F}_i, i=1,..4$ \cite{Chew:1957tf}.
The $\mathcal F_i$ are functions of the invariant mass $W$ and of the center-of-mass
(c.m.) scattering angle $\theta$. These four amplitudes determine fully the outcome
of any experiment. A determination of the four complex $\mathcal F_i$ amplitudes
obviously requires the measurement of at least seven different observables as functions of
$W$ and $\theta$, and one phase remains undetermined. A more detailed study shows that
such a model-independent amplitude analysis requires the measurement of at least
eight carefully chosen observables of
sufficient statistical accuracy~\cite{Sandorfi:2010uv,Ireland:2010bi}.

Recently, the CLAS collaboration reported precise data on the process $\gamma p\to
K^+\Lambda$. The differential cross section $d\sigma/d\Omega$ and the
$\Lambda$ recoil polarization $P$ were given in~\cite{McCracken:2009ra},
the polarization transfer from circular photon polarization to the $\Lambda$ hyperon
$C_x$ and $C_z$ in~\cite{Bradford:2006ba}, the beam asymmetry
$\Sigma$, the target asymmetry
$T$, and  the polarization transfer from linear photon polarization to the $\Lambda$ hyperon
$O_x$, $O_z$ were reported in~\cite{Paterson:2016vmc}. Whilst these represent eight measured
observables, data using a polarized target are still required to meet the
requirements for fully reconstructing the photoproduction amplitudes at each value of $W$ and $\theta$. Alternatively,
the angular dependence can be exploited, and the multipoles
can be fitted directly. This reduces the number of observables and the statistical precision of the data that are  required to get a fit~\cite{Wunderlich:2014xya}.

In this paper, we determine the multipoles driving
the process $\gamma p\to K^+\Lambda$ in 20\,MeV wide slices of  $K^+\Lambda$ invariant mass.
The formalism used to determine multipoles from data is described in \cite{Anisovich:2014yza}
where a first attempt was made to determine multipoles in slices of  invariant mass.
The observables are related to the $\mathcal F_i$ amplitudes; here we give one example.
The recoil polarization $P$ is given by
\be
 P\;I &=&
\sin(\theta) {\rm Im}[(2 {\mathcal F_2^*} + {\mathcal F_3^*}+ \cos(\theta) {\mathcal F_4^*}) {\mathcal F_1} +
                 \label{mult_1}\\
            && {\mathcal F_2^*}(\cos(\theta) {\mathcal F_3} + {\mathcal F_4}) +
                          \sin^2(\theta) {\mathcal F_3^*} {\mathcal F_4}]\,,\nonumber\ {\rm with}\\
    I &=& {\rm Re}[{\mathcal F_1} {\mathcal F_1^*} + {\mathcal F_2} {\mathcal F_2^*} -2 \cos(\theta) {\mathcal F_2} {\mathcal F_1^*} +  \frac{\sin^2(\theta)}{2} \times \nonumber \\&&\hspace{-4mm}
                ({\mathcal F_3} {\mathcal F_3^*} + {\mathcal F_4} {\mathcal F_4^*} + 2 {\mathcal F_4} {\mathcal F_1^*} +
              2 {\mathcal F_3} {\mathcal F_2^*} + 2 \cos(\theta) {\mathcal F_4} {\mathcal F_3^*})] \nonumber .
\ee


Once the $\mathcal F_i$ functions are known, they can be expanded into associated
Legendre functions $P_L(\cos\theta)$ and their derivatives $P^{\prime}_L(\cos\theta)$
with orbital angular momenta $L$ between the $K^+$ and $\Lambda$.
We have, e.g.,
\be \label{Eq:PW}
\hspace{-3mm}{\mathcal F_2}(W,\cos\theta) = \sum^{\infty}_{L=1} [(L+1)M_{L+}+LM_{L-}] P^{\prime}_{L}(\cos\theta).
\label{mult_2}
\ee
$E_{L\pm}$ and $M_{L\pm}$ are electric and magnetic multipoles
driving final states with defined orbital angular momentum $L$ between
meson and baryon and a total spin and parity $J^P=(L\pm 1/2)^\pm$.
Similar relations hold for the other three $\mathcal F_i$ functions \cite{Anisovich:2014yza}.

The number of multipoles increases considerably when higher orbital angular momenta are admitted,
and extremely precise data are required. Even then, for each slice in energy
and angle one phase remains undetermined. Hence one has to suppose that the phase of
one multipole amplitude is known which one might take from an energy-dependent fit. Clearly,
this introduces some model-dependence into the analysis.

\begin{figure}[pt]
\begin{tabular}{ccc}
\includegraphics[width=0.155\textwidth]{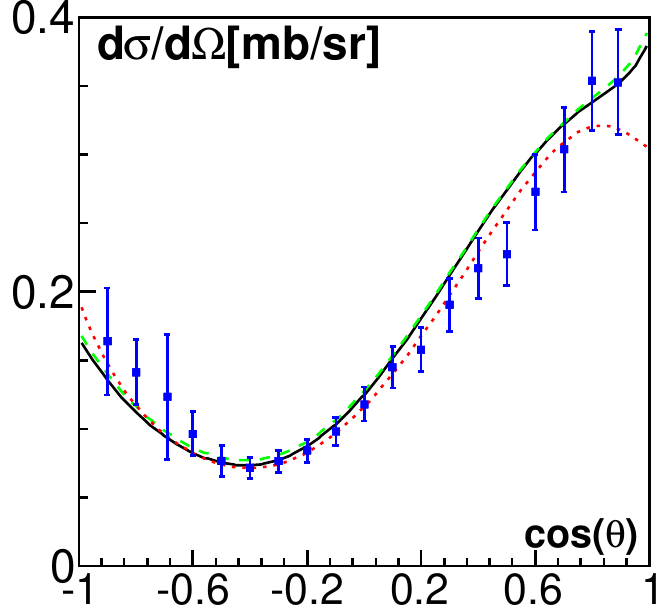}&
\hspace{-1mm}\includegraphics[width=0.155\textwidth]{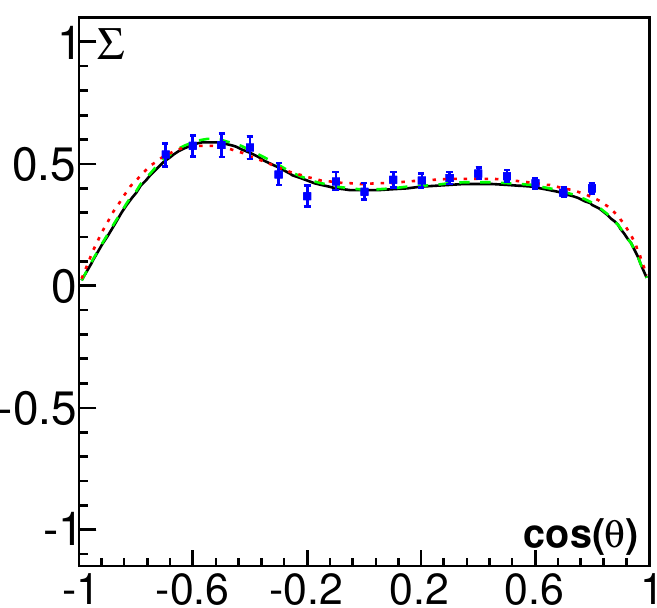}&
\hspace{-1mm}\includegraphics[width=0.155\textwidth]{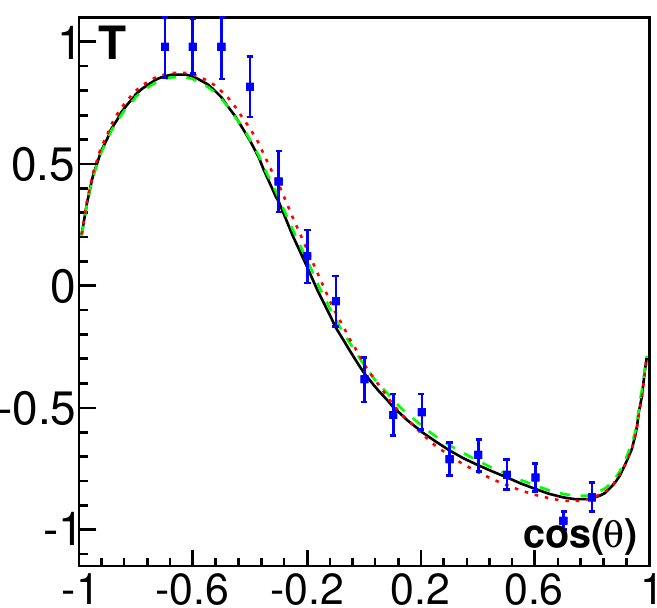}\vspace{-1mm}\\
\includegraphics[width=0.16\textwidth]{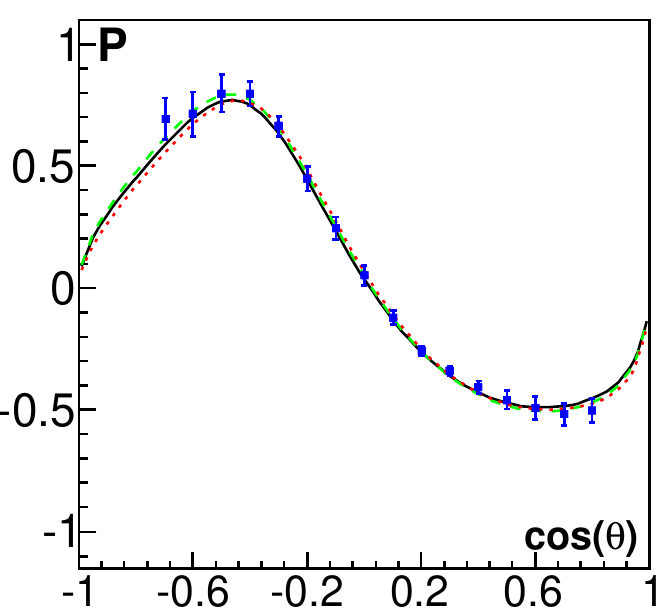}&
\hspace{-1mm}\includegraphics[width=0.155\textwidth]{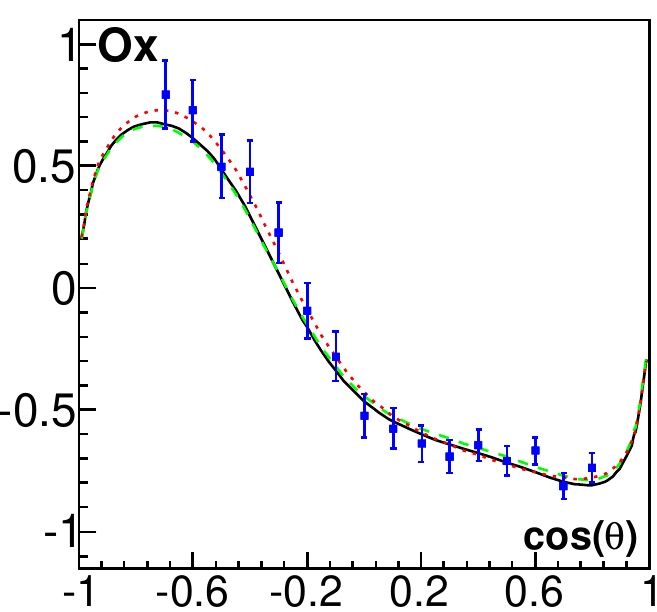}&
\hspace{-1mm}\includegraphics[width=0.155\textwidth]{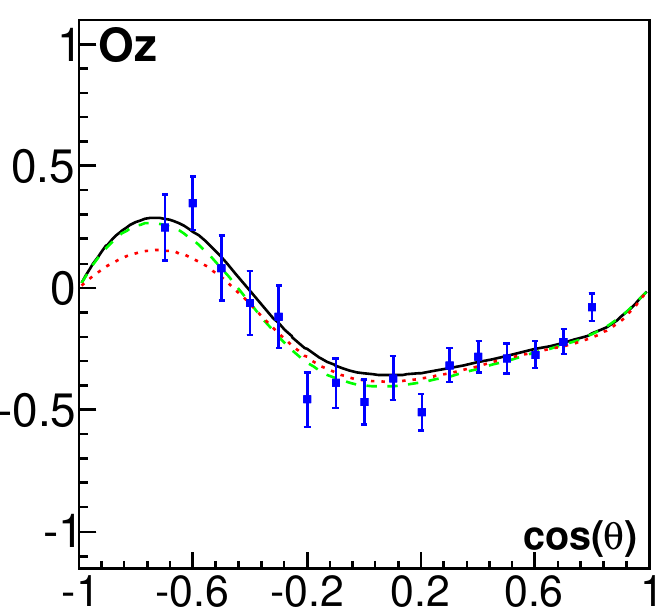}
\end{tabular}
\caption{\label{fig:data}(Color online) Example of a fit to the data for the mass range
from 1950 to 1970\,MeV. $d\sigma/d\Omega$:~\cite{McCracken:2009ra}, $P$~\cite{McCracken:2009ra},
$\Sigma$, $T$,
$O_x$, $O_z$~\cite{Paterson:2016vmc}. The (red) dotted curve corresponds to the fit used to
determine the multipoles of Fig.~\ref{fig:mult}, the (black) solid curves to fits using
$L+P$ for low-$L$ partial wave and BnGa for high-$L$, and the (green) dashed curves to the
BnGa fit. }
\end{figure}

Alternatively, the Legendre expansion of $\mathcal F_i$ functions (\ref{mult_2})
can be inserted into
the expressions for the polarization observables (\ref{mult_1}). In principle, this is an
infinite series that needs to be determined.
However, one can either truncate the power series at a given $L$, or one can
take the high-$L$ multipoles from a model. We use the high-$L$ multipoles from
a variety of solutions of the Bonn-Gatchina
(BnGa) fits~\cite{Anisovich:2015gia}.

Figure~\ref{fig:data} shows data on $\gamma p\to K^+\Lambda$ for one mass bin and
with three fit curves. The data on $C_x$, $C_z$, given in wider mass bins, are
mapped onto 20\,MeV bins and are used in addition. The red (dotted) curves in
Fig.~\ref{fig:data} show the result of a single-energy fit to the data.
With the given accuracy of the data, we found that
only a small number of multipoles -- $E_{0+}$,
$M_{1-}$, $E_{1+}$, $M_{1+}$ --  can be determined without imposing additional
constraints (like a penalty function which forces the fit
not to deviate too much from a predefined solution). The fit determines the real and imaginary parts of
these four photoproduction multipoles for one single mass bin.
These four multipoles varied freely in the fit, with no constraint. They excite resonances with the
quantum numbers $J^P=1/2^+$, $1/2^-$, and $3/2^+$. Three further multipoles -- $E_{2-}$,
$M_{2-}$, $E_{2+}$ driving excitations to $J^P=3/2^-$ and $5/2^-$ --  were constrained
to the energy-dependent BnGa fit by
a penalty function which forces the fit not to deviate too much from the predefined solution.
The higher mutipoles (up to $L<9$) were fixed to the energy-dependent BnGa fit. These
multipoles also provide the overall phase.

Figure~\ref{fig:data} shows two more fits: the solid curves represent the $L+P$ fit
(described below), the dashed curves the energy-dependent BnGa fit. The results
of the BnGa fits are shown in Table~\ref{tab:poles}. In the
fits, different BnGa starting fits were used which resulted from different fit hypotheses.
In particular, high-mass resonances with spin-parities $J^P=1/2^\pm, ...,7/2^\pm$ were added
to the fit hypothesis. The spread of the results was used to derive the errors given in Table~\ref{tab:poles}.

\begin{figure*}[pt]
\begin{center}
\begin{tabular}{cccc}
\hspace{-0.8mm}\includegraphics[width=0.245\textwidth,height=0.22\textwidth]{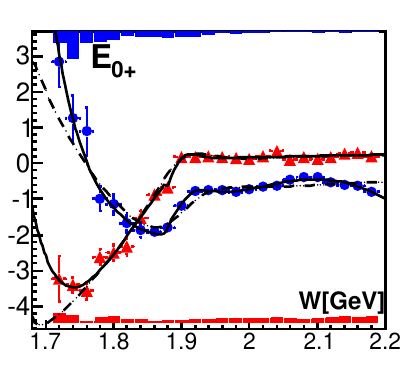}&
\hspace{1mm}\includegraphics[width=0.245\textwidth,height=0.22\textwidth]{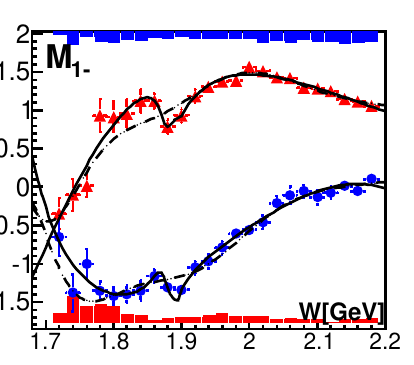}&
\hspace{0.8mm}\includegraphics[width=0.245\textwidth,height=0.22\textwidth]{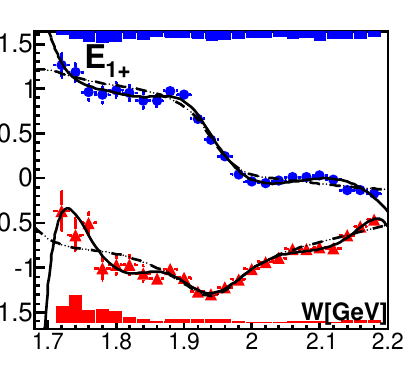}&
\hspace{0.8mm}\includegraphics[width=0.245\textwidth,height=0.22\textwidth]{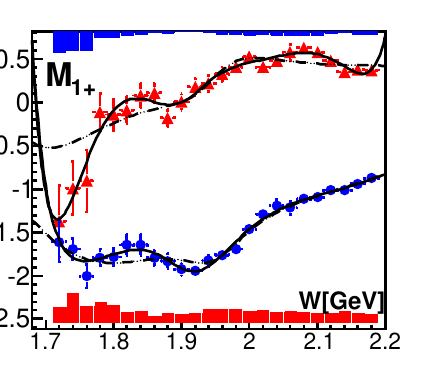}
\end{tabular}
\vspace{-4mm}
\end{center}
\caption{\label{fig:mult}(Color online) Real (red triangles) and imaginary (blue dots)
part of the $E_{0+}$, $M_{1-}$,  $E_{1+}$,
and $M_{1+}$ multipoles for the reaction $\gamma p\to K^+\Lambda$. The
systematic errors are given at the top (real part) and bottom (imaginary part)
of the subfigures. $E_{0+}$
excites the partial wave $J^P=1/2^-$; $M_{1-}$: $J^P=1/2^+$; $E_{1+}$
and $M_{1+}$: $J^P=3/2^+$. The solid curve shows the L+P fit, the dashed curve
the energy-dependent BnGa fit.\vspace{-3mm}
}
\end{figure*}

Figure~\ref{fig:mult} shows the multipoles as functions of the mass. The statistical
errors are determined by a scan of the $\chi^2$ dependence of the single-energy fit
on one of the multipoles while the other multipoles vary freely. The $\chi^2$ of this
fit includes the statistical and systematic errors of the data.
The systematic
errors for the real part are given at the top of the subfigures, those for the imaginary
part on the bottom. The systematic errors are determined by using different
energy-dependent BnGa fits, used to constrain the multipoles $E_{2-}$, $M_{2-}$, $E_{2+}$
and to determine the higher partial waves. The different energy-dependent BnGa fits
include, one by one, additional high-mass resonances (with weak evidence for
their existence) in each partial wave.
At small masses, there are visible differences between the $L+P$ fit and the BnGa fit.
These can be traced to the lack of polarization data at low energies in the backward
region.

First, we notice that all fitted multipoles show strong variations as functions of the mass. It therefore seems obvious that there are strong resonant contributions. Indeed, a first simple
fit with Breit-Wigner amplitudes plus a polynomial background shows that resonant contributions
are necessary for all four multipoles to achieve a good fit.

In this paper, we use a Laurent (more precisely Mittag-Leffler \cite{Mittag-Leffler})
method~\cite{L+P2013,L+P2013a,L+P2014,L+P2014a,L+P2015,L+P2016,L+P2016a}, called the L+P method,
to separate the singularities and the regular parts. The background is represented by analytic
functions with well defined cuts. The method was described by Ciulli and Fischer in \cite{CiulliFisher} and extensively used in the KH description of  $\pi N$
scattering~\cite{Hohler:1984ux} (details are described by Pietarinen in \cite{Pietarinen,Pietarinen1}).
The method is (almost) model independent. No dynamical assumptions are made except that
the scattering amplitude is an analytic function in the complex energy plane with singularities due to poles and thresholds.

The transition amplitude of the L+P model is parametrized as
\be \label{Eq:MCL+P}
T^a(W)&=&\hspace{-1mm}\sum _{j=1}^{{N}_{pole}} \frac{g^{a}_{j} }{W_j-W} \hspace{-0.5mm}
+\hspace{-0.5mm}\sum_{i=1}^3\sum_{k_i=0}^{K^{a}} c^{a}_{k_i}\left(\frac{\alpha^a_i\hspace{-1mm}-\hspace{-1mm}\sqrt{x^a_i-W}}{\alpha^a_i\hspace{-1mm}+\hspace{-1mm}\sqrt{x^a_i - W }}\right)^{k_i}\nonumber\\[-1ex]
\ee
where $a$ is a channel index, $W_j$ are pole positions in the complex $W$ (energy) plane, $g^{a}_j$
are residues for $\pi N\to K\Lambda$ transitions. The $x^{a}_i$ define the branch points,
$c^{a}_{k_i}$, and $\alpha^{a}_i$ are real coefficients.
$K^a, \, L^a, \, M^a \, ... $\, are the number of Pietarinen coefficients in channel $a$.
The first part represents the poles
and the second term three branch points. The first branch point is chosen to describe all
subthreshold and left-hand cut processes, the second one is
fixed to the dominant channel opening, and the
third one  represents background contributions
of all channel openings in the physical range.

To enable the fitting we define a reduced discrepancy function $D_{dp}$ as:
\be
\label{eq:Laurent-Pietarinen}
 D_{dp} &=&\sum _{a}^{all}D_{dp}^a;\qquad  D_{dp}^a =  \frac{1}{2 \, N_{W}^a - N_{par}^a} \times    \nonumber \\
& &\hspace{-5mm}\sum_{i=1}^{N_{W}^a}
    \left\{ \left[ \frac{{\rm Re} \,T^{a}(W^{(i)})-{\rm Re} \, T^{a,exp}(W^{(i)})}{ Err_{i,a}^{\rm Re}}  \right]^2 \right.\nonumber\\
&&\left.\hspace{-5mm}   +\quad\ \,\left[ \frac{{\rm Im} \, T^{a}(W^{(i)})-{\rm Im} \, T^{a,exp}(W^{(i)})}{ Err_{i,a}^{\rm Im}} \right]^2 \right\} + {\cal P}^a \nonumber\\
    {\rm where} \ && {\cal P}^{a} = \lambda^a_{k_1} \sum _{k_1=1}^{K^a} (c^a_{k_1})^2 \, {k_1}^3 +   \nonumber \\
&+& \lambda_{k_2}^a \sum _{k_2=1}^{L^a} (c^a_{k_2})^2 \, {k_2}^3 + \lambda_{k_3}^a \sum _{m=1}^{M^a} (c^a_{k_3})^2 \, {k_3}^3 \nonumber
 \ee
 is the Pietarinen penalty function which ensures fast and optimal convergence.

 $N_{W}^a$ is the number of energies in channel $a$, $N_{par}^a$ the number of fit parameters in channel $a$,
 $\lambda_c^a, \lambda_d^a, \lambda_e^a$ are Pietarinen weighting factors,
$Err_{i,a}^{\rm Re, \, Im} ..... $  errors of the real and imaginary part, and
$c_{k_1}^a, c_{k_2}^a, c_{k_3}^a$ real coupling constants.

Figure~\ref{fig:mult} shows the $L+P$ fit. This fit follows the ``data'' much more precisely than the BnGa fit. The reason is, of course, that BnGa fits the real data while $L+P$ represents
a fit to the ``data'' in Fig.~\ref{fig:mult}.

In addition to the data on $\gamma p\to K^+\Lambda$ discussed here
(channel $a=1$), we included in the fits the amplitudes $T_{J^P}$ for $\pi^-p\to K^0\Lambda$
(channel $a=2$)~\cite{Anisovich:2013tij} since they provide the information on the $\pi N\to K\Lambda$  transition residues and allowed for a better determination of the low-mass poles
at the $K\Lambda$ threshold. The photoproduction data alone determine well the
properties of the resonance at about 1900\,MeV but not the poles at the $K\Lambda$
threshold. The $N(1720)3/2^+$ resonance cannot be determined reliably from the
$K\Lambda$ final states.

 \begin{table*}[ht]
\caption{\label{tab:poles}  Properties of nucleon resonances from
the Particle Data Group (PDG estimates) \cite{Olive:2016xmw}, the BnGa PWA fit,
and from $L+P$ fits. Masses and widths are given in MeV, the normalized inelastic pole
residues $2\cdot g^{a}(\pi N\to K\Lambda)/\Gamma_{a}$ are numbers.
}
\def\arraystretch{1.2}
\begin{tabular}{c|ccc|ccc|ccc}  \hline
                         & \multicolumn{3}{c|}{$J^P=1/2^-$}
                         & \multicolumn{3}{c|}{$J^P=1/2^+$}
                         & \multicolumn{3}{c}{$J^P=3/2^+$}         \tabularnewline\hline
                         &  PDG             &    BnGa     &  ${\mathrm MC \, \, L+P}$
                           &  PDG             &    BnGa     &  ${\mathrm MC \, \, L+P}$
                         &  PDG             &    BnGa     &    L+P    \tabularnewline\hline
 M$_1$                   &  1640-1670       & $1658\pm10$  &   ${1660 \pm 5}$
                         &  1670-1770       & $1690\pm15$ &   ${1697 \pm 23}$
                                               &   -    & - &         -        \tabularnewline
  $\Gamma_1$             &   100-170        & $102\pm8$   &     ${59 \pm 16}$
                         &    90-380        & $155\pm25$  &    $84\pm 34$
                                                 &   -       & - &          -       \tabularnewline
$|Res_{_1}(\pi N$$\to$$K\Lambda)|$&         &$0.26\pm0.10$&   ${0.10 \pm 0.10}$
                           &  -               &$0.16\pm0.05$&${0.12^{+0.24}_{-0.12}}$
                                                 &   -              &-&          -        \tabularnewline
    $\Theta_1$           &  -               &$(110\pm20)^0$&    ${(95 \pm 33)^0}$
                             &     -            &$-(160\pm25)^0$&   ${-(119 \pm 83)^0}$
                                                 &    -             &-&          -        \tabularnewline\hline
 M$_2$                   &      -           &$1895\pm15$  &    ${1906 \pm 17}$
                         &      -           &$1860\pm40$  &    ${1875 \pm 11}$
                                                 &   1900-1940      &$1945\pm35$  &   $1912 \pm 30 $   \tabularnewline
  $\Gamma_2$             &      -           &$132\pm 30$  &   ${100 \pm 10}$
                           &      -           &$230\pm50$    &   ${33 \pm 9}$
                                                 &   130-300        &$135^{+70}_{-30}$&$  166 \pm 30$   \tabularnewline
  $|Res_{_2}(\pi N$$\to$$K\Lambda)|$&  -    &$0.09\pm0.03$ &    ${0.06 \pm 0.02}$
                           &      -           & $0.05\pm0.02$& ${0.30 \pm 0.10}$
                                                 &      -           & $0.03\pm0.02$&  $  -$    \tabularnewline
    $\Theta_2$           &      -           &$(8\pm30)^0$&    ${(87 \pm 27)^0}$
                             &      -           &$(27\pm30)^0$ &   ${(82 \pm 9)^0}$
                                               &      -           &$(90\pm40)^0$&  $   -$ \tabularnewline\hline
 \end{tabular}
\end{table*}

\paragraph{\bf\boldmath $E_{0+}$ and $T_{1/2^-}$:}
First, the branch points were fixed to the $\pi N$, $K^+\Lambda$, and the $\eta^\prime p$ thresholds.
A fit with one pole failed to reproduce both channels, and a second pole was added. The fit gave
a reasonable description of the data and was slightly improved when the second and third branch
points were released to adjust to close-by values. The results of the fit are
given in Table~\ref{tab:poles}. The agreement between the BnGa and the $L+P$ fit is excellent.
We consider the existence of both resonances as certain
and its properties as reasonably well defined.

\paragraph{\bf\boldmath $M_{1-}$ and $T_{1/2^+}$:}
The procedure was repeated for the $J^P=1/2^+$ partial wave. Again, the fits requires two
resonances, in particular the new $N(1880)1/2^+$ state, but it turns out to be rather narrow.
The second branch point was fixed at the $K^+\Lambda$ threshold, the third one moved
to $x_3=1.898$~GeV. The result of the $L+P$ and the BnGa fits are given in Table~\ref{tab:poles}.
The existence of both resonances is mandatory in both fits. However, there is a discrepancy in
the width. We changed the binning by shifting the bins by 10\,MeV and by choosing 25\,MeV
bins; the narrow structure in the $L+P$ fit remained. The narrow width is however incompatible
with the BnGa fit: when a $N(1880)1/2^+$ width of 42\,MeV was imposed,  the overall
$\chi^2$ deteriorated by 1000 units, and the fit visibly missed to describe the data properly.

When a $N(1880)1/2^+$ width of 150\,MeV was imposed in the $L+P$ fit, it deteriorated
from $\chi^2=16.7$ for 76 data points and 43 parameters to $\chi^2=24.9$ for 40 parameters.
Compared to the actual data, the difference of the main $L+P$ fit (with 33\,MeV width) and the
test fit (with 150\,MeV width) is marginal. For the 674 data points, the improvement
in $\chi^2$ is 4.5 only. This gain does not justify to claim a narrow structure
in the $M_{1-}$ multipole. It seems that, in the energy independent analysis, small
systematic deviations from the ``true'' values create a narrow structure which can
interpreted as a narrow resonance. The existence of $N(1880)1/2^+$ is certain but
the $N(1880)1/2^+$ width is not well defined, likely due to a statistical fluctuation
(or a systematic deviation) in one of the data sets.

\paragraph{\bf\boldmath $E_{1+}$ and $M_{1+}$:}
 In this case, only single-channel analyses were performed as the data from \mbox{$\pi^-p\to K^0\Lambda$} process were of insufficient quality. In particular, the properties of $N(1720)3/2^+$ could
not be deduced from the fits. The $E_{1+}$, $M_{1+}$ multipoles were fitted
simultaneously with identical pole positions, the same branch points but with free Pietarinen
coefficients. The second branch-point was fixed to the $K\Lambda$ threshold, the third branch-point
converged to $x_3=2.46$ GeV. The results of the fit are reproduced in Table~\ref{tab:poles}.
We consider the existence of the $N(1900)3/2^+$ resonance as certain
and its properties as reasonably well defined.

The use of two independent approaches, one energy-independent and one energy-dependent
allowed us to draw definitive conclusions about the existence of several excited
nucleon states. So far, the evidence for two of these resonances was estimated
by the PDG
{\it to be fair only}. These two resonances, $N(1880)1/2^+$ and $N(1895)1/2^-$,
are presently not included in the PDG Baryon Summary Table and are mostly not
taken into account when models of baryons are compared to data. Establishing the
existence of nucleon resonances in this mass range is therefore of great importance.

Summarizing, we have determined low-$L$ multipoles for the reaction
$\gamma p\to K^+\Lambda$. The multipoles were fitted using the Laurent-Pietarinen
method, which has minimal model dependence. The fits firmly establish the existence of
three resonances and determined their properties. This opens up a promising new
avenue of the field of baryon spectroscopy with electromagnetic probes.

\begin{acknowledgments}
This work is supported by the \textit{Deutsche Forschungsgemeinschaft} (SFB/TR110, SFB1044),
the \textit{U.S. Department of Energy, U.S. National Science Foundation, the UK Science and Technology Facilities Council (ST/L005719/1), and the
\textit{Russian Science Foundation} (RSF 16-12-10267).}
\end{acknowledgments}

\end{document}